\documentclass[11pt,letterpaper]{JHEP3}
\usepackage{epsfig}
\def\fin{{f_\infty}}
\def\lp{\ell_P}

\newcommand{\be}{\begin{equation}}
\newcommand{\ee}{\end{equation}}
\newcommand{\bea}{\begin{eqnarray}}
\newcommand{\eea}{\end{eqnarray}}
\newcommand{\ba}{\begin{eqnarray}}
\newcommand{\ea}{\end{eqnarray}}

\newcommand{\beq}{\begin{equation}}
\newcommand{\eeq}{\end{equation}}
\newcommand{\beqa}{\begin{eqnarray}}
\newcommand{\eeqa}{\end{eqnarray}}
\newcommand{\beqar}{\begin{eqnarray*}}
\newcommand{\eeqar}{\end{eqnarray*}}

\def\R{{\mathcal R}}
\def\G{{\mathcal G}}
\title{On the new massive gravity and AdS/CFT}
\author{Aninda Sinha \\
\it Perimeter Institute for Theoretical Physics\\
\it Waterloo, Ontario N2L 2Y5, Canada\\
\vskip .5cm

{\rm E-mail:}\ \ {\tt asinha@perimeterinstitute.ca}}

\abstract{Demanding the existence of a simple holographic $c$-theorem, it is shown that a general (parity preserving) theory of
gravity in 2+1 dimensions involving upto four derivative curvature invariants reduces to the new massive
gravity theory. We consider extending the theory including upto six derivative curvature
invariants. Black hole solutions are presented and consistency with 1+1 CFTs is checked. We present evidence that
bulk unitarity is still in conflict with a positive CFT central charge for generic choice of parameters.  However, for a special choice of parameters appearing in the four and six derivative terms reduces the linearized equations to be two derivative, thereby ameliorating the unitarity problem. \vskip 1cm {\it \hskip 4cm Dedicated to the loving memory of my father}}


\preprint{arXiv:1003.0683 [hep-th]}
\begin{document}
%\maketitle
%\vskip 5cm

%\newpage

\tableofcontents
\section{Introduction}
 Recently a higher derivative theory of gravity has been proposed \cite{nmg} in 2+1 dimensions which has nice properties\cite{nmg2,nmg3,many,unit}; it
 is parity preserving and unitary--it has been christened ``new massive gravity (NMG)". In asymptotically Minkowski spacetime the theory is equivalent to a
 Pauli-Fierz action for a massive spin-2 field at the linearized level. The NMG theory has also been considered in the
 context of the AdS/CFT correspondence\cite{mgads,mgads2,mgads3}. It was found that bulk unitarity for the massive spin-2 modes implied negative central charge in the CFT.

 This theory is interesting for a different reason. In \cite{gbhd,gbhd2,gbhd3}, in the context of holography,  various properties of Gauss-Bonnet
 gravity in arbitrary spacetime dimensions $D\geq 5$ were studied. Since in 2+1 dimensions, Gauss-Bonnet vanishes identically, a natural question arises: What is the analogous theory in lower dimensions? In a certain sense, we
 will argue in this paper that this is nothing but the NMG! The equivalence arises as follows: what
 characterises the dual CFTs are central charges which in even spacetime dimensions arise as conformal
 anomalies. It is shown in \cite{ms} that for these GB theories in higher dimensions, there exists a
 ``c"-theorem, where the anomaly associated with the Euler density admits for a monotonic function defined along an
 RG-flow induced by matter satisfying the null energy condition \cite{cth}. We will show that the NMG theory admits a
 ``c"-function which is in conjunction with the $c$-theorem in 1+1 dimensional CFTs. The UV value of the anomaly
 is precisely what is predicted in \cite{ms}. In \cite{ms} it is shown that the anomaly associated with the Euler density for Gauss-Bonnet gravity with action
 \be\label{gend} I=\frac{1}{2\lp^{D-2}}\int d^D x \sqrt{-g}(R+\frac{(D-1)(D-2)}{L^2}+\frac{\lambda
L^2}{(D-3)(D-4)} [R_{abcd}R^{abcd}-4 R_{ab}R^{ab}+R^2])\,,\ee
 in an arbitrary dimension is given by
 \be a_D=\pi^2\frac{\tilde L^{D-2}}{\lp^{D-2}}(1-2\frac{D-2}{D-4}\lambda \fin)\,, \ee
where $D$ is odd so that the CFT dual is in even dimensions, $\tilde L$ is the AdS radius and $\fin$ defined through $\tilde L=L/\sqrt{\fin}$ satisfies $1-\fin+\lambda f^2_\infty=0$. While this formula is strictly valid for $D\geq 5$, we can extrapolate to $D<5$. Setting $D=3$, we find that the anomaly coefficient is proportional to $\tilde L/\lp (1+2\lambda \fin)$. We will find (once the correct normalizations are chosen) that this is precisely what the NMG theory yields \footnote{A higher dimensional generalization of the NMG was attempted in \cite{no} but the theory was found to be non-unitary even around flat space. }.

 We will examine higher derivative gravity in 2+1 dimensions in the context of the AdS/CFT correspondence. Since
 there is a $c$-theorem in 1+1 dimensional CFT's, one wonders if somehow demanding the existence of
 such a theorem holographically will be of any redeeming value\footnote{In the context of topologically massive gravity of \cite{tmg}, holographic RG flows were studied in \cite{tmgcth}.}; for instance can it constrain the parameters entering the gravity lagrangian? We start with an
 arbitrary four-derivative theory in 2+1 dimensions in the presence of a negative cosmological constant. As we
 will show, the combination of higher derivative terms that appear in NMG allows for such a $c$-theorem. Conversely, we will show that the NMG is the only choice that allows for a $c$-theorem of analogous form. Inspired by
this, we consider adding six derivative terms and derive constraints on the parameters which allows for a natural
 extension of such a $c$-theorem. Black hole solutions are presented and their thermodynamics considered. The
 resulting entropy of such black holes are consistent with known results in 1+1 CFTs, namely that it is
 proportional to the central charge times temperature. We should note here a crucial difference between the NMG
 and GB gravity in higher dimensions. The equations of motion for GB gravity is second order but for the NMG
 theory they are intrinsically fourth order. In \cite{mr} a six derivative theory was constructed in $D\geq 5$, where the
 equations of motion with only radial dependence was shown to be second order. Our six derivative theory is the
 lower dimensional cousin of this theory.  In our case, we will find that as in the NMG theory, the extended NMG allows for a
 certain factorisation of the operator arising in the equations of motion describing linearization about the AdS
 vacua\cite{mgads2}. The equations of motion quite generally involve fourth order in derivatives.

 The fact that the equations of motion involve higher derivatives leads to the question of unitarity. Around flat
 space, the NMG theory is known to be unitary if the wrong sign kinetic term is taken. In \cite{mgads} it was
 shown that in AdS bulk unitarity was in conflict with the positivity of a CFT central charge. This is of course in contrast to what happens in GB holography.
 We analyse the issue of unitarity in this extended version of NMG. We present evidence that the
 conflict between bulk unitarity and positivity of the CFT central charge still persists in the six derivative
 theory.

 The paper is organized as follows. In section 2, we consider holographic $c$-theorems in higher derivative
 gravity in 2+1 dimensions. We show that at the four derivative level, the NMG naturally admits a $c$-function. We construct
 a new theory at the six derivative order which admits such a $c$-function.
 In section 3, we consider the Weyl anomaly in the dual CFT and show that it is equivalent to the UV value of the $c$-function.
 In section 4, we present black hole solutions to the six derivative theory and show that the entropy is
 consistent with what is expected from a 1+1 CFT. In section 5, we turn to the issue of unitarity in the bulk vs
 positivity of the CFT central charge. First we show that the radial part of the equations of motion for the
 bulk mode admits a factorization similar to that in the four derivative theory. Using this we argue
 that the conflict between bulk unitarity and positivity of the CFT central charge persists in this new theory.
 We conclude in section 6. Appendix A contains a discussion about how to extend our construction further to eight derivative
 order which admits a Pauli-Fierz type lagrangian. Appendix B has a discussion of Lifshitz solutions. Appendix C contains more details about the $c$-function in NMG.

\section{Holographic $c$-theorems and a new derivation of NMG}
We wish to see whether demanding the existence of a holographic $c$-theorem will give rise to any interesting constraint. The action we will consider is
\be \label{actmain}
I=\frac{1}{2\lp}\int d^3 x \sqrt{-g}(R+\frac{2}{L^2}+L^2 {\mathcal R}_2+L^4 {\mathcal R}_3)\equiv \frac{1}{2\lp}\int d^3 x \sqrt{-g}(R+\frac{2}{L^2}+K) \,,
\ee
where
\begin{eqnarray}\label{lagmain}
{\mathcal R}_2&=& 4(\lambda_1 R_{a b}R^{a b}+\lambda_2 R^2)\,,\\
{\mathcal R}_3 &=&\frac{17}{12}(\mu_1 R_a^b R_b^c R_c^a+\mu_2 R_{ab}R^{ab}R+\mu_3 R^3
+\mu_4 \nabla_a R_{bc}\nabla^a R^{bc} +\mu_5 \nabla_a R \nabla^a R)\,.
\end{eqnarray}
We will set $\mu_4=\mu_5=0$ to get rid of terms in the equations of motion having six derivatives\footnote{This is a choice and one could work keeping these terms. For a $c$ function, these terms will have to be set to zero as we will argue shortly.}. The reason for having the factors $4$ and $17/12$ will become clear shortly. The equations of motion for this action are given by
\be
R_{ab}-\frac{1}{2}R g_{ab}-\frac{1}{L^2}g_{ab}-K_{ab}=0\,,
\ee
where
\begin{eqnarray}
K_{ab} &=& 4 L^2\bigg{(} \lambda_2(-2 R R_{ab}+2 \nabla_a \nabla_b R+ g_{ab}[\frac{1}{2} R^2-2 \nabla^2 R])\nonumber \\
    &&~~~~~~~~ +\lambda_1 (-2 R_a^c R_{cb}+2 \nabla_c \nabla_{(a} R_{b)}^c-\nabla^2 R_{ab}+g_{ab}[\frac{1}{2} R_{cd}R^{cd}-\frac{1}{2}g_{ab}\nabla^2 R])\bigg{)}\nonumber \\
    &+& \frac{17}{12}L^4 \bigg{(}\mu_1(-3 R_{ac}R^c_d R^d_b-\frac{3}{2} \nabla^2 R_{ac}R^c_b +3 \nabla_c\nabla_{(a} R_{b)}^d R_d^c+g_{ab}[\frac{1}{2}R_c^d R_d^e R_e^c-\frac{3}{2} \nabla_d \nabla_c R^c_e R^{ed}])\nonumber \\
    &&~~~~~~~~~\mu_2(-R_c^d R_d^c R_{ab}-2 R R_{ac}R^c_b +\nabla_b \nabla_a R_{cd}R^{cd}-\nabla^2 R R_{ab}+2 \nabla_c \nabla_{(b}R_{a)}^c R\nonumber \\ && ~~~~~~~~~~~~~~~~~~~~~~~+g_{ab}[\frac{1}{2} R R_{cd}R^{cd}-\nabla_d\nabla_c R R^{cd}-\nabla^2 R_{cd}R^{cd}])\nonumber \\
    &&~~~~~~~~~\mu_3 (-3 R^2 R_{ab}+3 \nabla_a \nabla_b R^2+g_{ab}[\frac{1}{2}R^3-3 \nabla^2 R^2])\bigg{)}\,.
\end{eqnarray}
In the absence of the six derivative term, $K_{ab} g^{ab}=-K/2$ but this is no longer true in the presence of the six derivative term\footnote{However, if we linearize around flat space, this will not spoil the unitarity and renormalizability of the four derivative theory since around flat space $R_{ab}\sim 0$.}. Quite generally, the above lagrangian will admit for Lifshitz solutions which are discussed in appendix B. We will focus here on AdS solutions.

Let us now examine the possibility of a holographic $c$-theorem following \cite{cth}. This requires the addition of a matter sector
which will induce a RG-flow. In Einstein gravity this works as follows. We begin with the ansatz
\be\label{ans1}
ds^2=e^{2A(r)} (-dt^2+dx^2)+dr^2\,.
\ee
The equations of motion imply
\be
A''=T^t_t-T^r_r \leq 0\,,
\ee
where $T_{ab}$ is the stress-energy tensor associated with the matter. We have assumed that the matter sector satisfies the null-energy condition, i.e., $T_{ab}\zeta^a \zeta^b \geq 0$ for any arbitrary null vector $\zeta$. Then choosing $\zeta^a\equiv \{\zeta^t,\zeta^r,\zeta^x\}=v\{e^{-A},1,0\}$ leads to $T_{tt}e^{-2A}+T_{rr}\geq 0$ or $T_t^t-T_r^r\leq 0$. Define
\be \label{clead}
c(r)=\frac{1}{\lp A'(r)} \,,
\ee
so that
\be
c'(r)=-\frac{A''}{\lp A'^2}\geq 0\,.
\ee
Thus $c(r)$ is monotonically increasing. In the absence of matter, $A(r)=r/L$ so that the UV boundary is $r\rightarrow \infty$ for which $c(r)$ becomes the central charge of the dual CFT. In \cite{cth} an example was shown where a kink solution existed interpolating two different solutions which asymptotically had $A(r)$ linear with different slopes. If we assume that spacetime asymptotes to AdS both in the UV and IR, then identifying $r\rightarrow \infty$ as the UV and $r\rightarrow -\infty$
as the IR \cite{cth} in these coordinates we thus have $c_{UV}\geq c_{IR}$. Now let us turn to the case where we have added
${\mathcal R}_2$. Here we find
\begin{eqnarray}
T^t_t-T^r_r&=&
A''[1-16 A'' L^2(8 \lambda_2+3\lambda_1)+16 A'^2 L^2(3 \lambda_2+\lambda_1)]-8 L^2(8 \lambda_2+3\lambda_1)A'
A'''\nonumber\\&&- 4 L^2(8\lambda_2+3\lambda_1) A''''  \,.
\end{eqnarray}
We want to find a function that generalizes (\ref{clead}). We will demand that it is made of $A'$ and derivatives thereof. Following the above logic, taking the derivative of this function will be proportional to the equations of motion. In the GB case in higher dimensions, the form for the $c$-function was \cite{ms} $c\sim (1+\alpha \lambda L^2 A'^2)/A'$ with $\alpha=const$. If we demand the same simple form for the $c$-function as in the GB case, it turns out that only way of accomplishing this is to set the four derivative terms in the equations of motion for $A$ to zero. In other words, we want the form of  $T_t^t-T_r^r$ to be a simple one depending only on $A''$ and $A'$ and no further higher derivative terms involving e.g. $A'''$ or $A''''$.
Setting $\lambda_2=-3/8\lambda_1$ gets rid of the four derivative terms and leads to
\be
A''(1-2 A'^2 L^2 \lambda_1)=T^t_t-T^r_r\,.
\ee
Thus defining
\be
c(r)=\frac{1}{\lp A'}(1+2\lambda_1 L^2 A'^2)\,,
\ee
leads to
\be \label{only}
c'(r)=-\frac{T^t_t-T^r_r}{\lp A'^2}\geq 0\,.
\ee
A discussion, showing that the only way to get a $c(r)$ satisfying eq.(\ref{only}) is to choose $\lambda_2=-3/8\lambda_1$, is given in appendix C. Note that with $\lambda_2=-3/8\lambda_1$ , the ${\mathcal R}_2$ term is exactly what appears in the NMG theory! Hence
this is another way of getting the NMG theory. In a similar manner we can show that the choice for ${\mathcal R}_3$
\be
\mu_1=\frac{64}{17}\mu_3\,,\quad \mu_2=-\frac{72}{17}\mu_3\,,\quad \mu_{4,5}=0\,,
\ee
will lead to
\be
A''(1-2 A'^2 L^2 \lambda_1-3 A'^4 L^4 \mu_3)=T^t_t-T^r_r\,.
\ee
Then defining
\be
c(r)=\frac{1}{\lp A'}(1+2\lambda_1 L^2 A'^2+\mu_3 L^4 A'^4)\,,
\ee
will satisfy
\be
c'(r)\geq 0\,.
\ee
Hence using this theory will admit a function $c(r)$ which is monotonically increasing. We will call this theory the extended NMG \footnote{Curiously, after my paper was released, it was realized in \cite{tekin} that the combination of terms appearing in the NMG and the extended version appear also in a DBI version expanded upto $O(R^3)$.}. We should also point out that we have demonstrated the existence of a {\it certain} $c$-function. There could be more such functions exhibiting monotonicity whose physical interpretation may be different. Our $c$-function is the one which naturally leads to the central charge $c$ characterizing the 1+1d CFTs.
In the absence of a matter sector
\be \label{norma}
A(r)=\frac{r}{\tilde L}\equiv \frac{r \fin^{1/2}}{L}\,, \qquad 1-\fin+f^2_\infty \lambda_1 +f^3_\infty \mu_3=0\,,
\ee
satisfies the equations of motion\footnote{Notice the following. Suppose we did not have an explicit cosmological constant in eq. (\ref{actmain}). This would simply lead to dropping the 1 in the second equation in (\ref{norma}). Without higher derivative corrections we would have $\fin=0$ and no AdS. However, with higher derivative corrections $\fin\neq 0$ is a solution and it is easy to have $\fin>0$ leading to AdS. Even in this case, all the formulae and the analysis in what follows go through.}. This explains the choice of the normalization in (\ref{lagmain}) and is in conjunction with \cite{gbhd, mr}. For this
\be \label{csix}
c=\frac{\tilde L}{\lp}(1+2\lambda_1 \fin +\mu_3 f^2_\infty)\,,
\ee
which will be identified as the Weyl anomaly\footnote{The CFT central charge frequently quoted in the literature is $12\pi$ times $c$.}in the next section. The Weyl anomaly for the six-derivative theory in five dimensions \cite{mr} was calculated in \cite{mps}. Eq. (\ref{csix}) is the $D=3$ extrapolation of the anomaly associated with the Euler density in the six-derivative theory in higher dimensions. It thus appears that once we fix the normalization such that eq. (\ref{norma}) is satisfied, the (extended) NMG is the ``natural" cousin of the higher dimensional (six-derivative extended) Gauss-Bonnet theory.
\section{Weyl anomaly and the central charge $c$}
The easiest way to derive the Weyl anomaly coefficient \cite{ejm} of the CFT is to consider the theory on an $S^2$ with the
metric
\be
ds^2=\frac{dr^2}{1+f(r)}+g(r)(d\theta^2+\sin^2\theta d\phi^2)\,,
\ee
with
\be
f(r)=\frac{r^2}{\tilde L^2}\,,\qquad g(r)=r^2\,,
\ee
which solve the equations of motion.
Then the action evaluates to
\be
I=-\frac{8\pi}{\lp \tilde L} (1+2\fin\lambda_1 +f^2_\infty \mu_3)\int_0^\Lambda dr \frac{r^2}{\sqrt{r^2+\tilde
L^2}}\,,
\ee
where $\Lambda$ is a cutoff we have put by hand. Evaluating the integral leads to
\be
I=-\frac{8\pi}{\lp \tilde L} (1+2\fin\lambda_1 +f^2_\infty \mu_3)\left(\frac{\Lambda^2}{2}-\frac{\tilde L^2}{2} \log
\frac{2\Lambda}{\tilde L}\right)\,.
\ee
Here we expect the quadratic divergence to be cancelled when one takes into account the surface term and the
appropriate counterterm. The coefficient of the log term is associated with the Weyl anomaly. This can be
written as
\be
c_{log}= \frac{4 \pi \tilde L}{\lp} (1+2\fin \lambda_1 +f^2_\infty \mu_3)= 4\pi c\,.
\ee
Thus we find that the $c$-theorem argument of the previous section leads to precisely the Weyl anomaly coefficient
(upto an irrelevant factor of $4\pi$).

\section{Black hole solutions}
For the extended NMG theory in AdS, one can easily find black hole solutions which are similar to the black holes in the NMG \cite{mgads,Oliva:2009ip}. We make the ansatz
\be\label{bhmet}
ds^2=-N(r)^2 f(r)\frac{r^2}{L^2}dt^2+L^2\frac{dr^2}{r^2 f(r)}+\frac{r^2}{L^2}dx^2\,.
\ee
Then it can be shown that
\be\label{bhmetp}
N(r)=const\,,\qquad f(r)=\fin+\frac{c_1}{r}+\frac{c_2}{r^2}\,,
\ee
will lead to
\be
\frac{c_1}{L^3}(-1+2\fin \lambda_1 +3f^2_\infty \mu_3)=0\,,
\ee
from which we see that either $c_1=0$ with $(-1+2\fin \lambda_1 +3f^2_\infty \mu_3)\neq 0$. If $(-1+2\fin \lambda_1 +3f^2_\infty
\mu_3)=0$, then $c_1$ need not be zero and an extremal black hole solution is possible. We will not consider
this interesting possibility any further and will focus on the case where $c_1=0$. We will further set
$N(r)^2=1/\fin$ so that the velocity of light in the CFT is unity.
\subsection{Thermodynamics}
We will choose
\be
c_2=-\fin r_0^2\,,
\ee
so that the horizon is at $r=r_0$.
Since we have not worked out the generalized Gibbons-Hawking term or the requisite counterterms, we will use
background subtraction to examine the thermodynamics of these black holes. We will use the empty AdS as the
background to subtract. The temperature is given by
\be
T=\frac{r_0}{\pi L \tilde L}\,.
\ee
This will lead to the free energy density to be
\be
{\mathcal F}=\frac{T}{V}(I_E-I_E^{0})=-\frac{\pi^2 T^2 \tilde L}{\lp}(1+2\fin \lambda_1+f^2_\infty \mu_3)\,.
\ee
Using this we calculate the entropy density to be
\be\label{ent}
s=-\frac{\partial {\mathcal F}}{\partial T}=\frac{2\pi^2 T \tilde L}{\lp}(1+2\fin \lambda_1+f^2_\infty
\mu_3)=2 \pi^2  c T\,,
\ee
which is exactly what we expect for a 1+1 CFT.
\subsection{Entropy using the Wald formula}
Equation (\ref{ent}) can be verified using Wald's formula as well. Recall Wald's formula for entropy \cite{wald}
\be
S=-2\pi \oint dx \sqrt{h} \frac{\partial {\mathcal L}}{\partial R_{abcd}} \hat \epsilon_{ab} \hat \epsilon_{cd}\,,
\ee
where $\hat \epsilon_{ab}$ is the binormal to the horizon. For the action in eq. (\ref{actmain}), this works out to be
\begin{eqnarray}\label{wald}
S=\frac{2\pi A}{\lp}\bigg{(}1&+&8\lambda_2 L^2 R+4 \lambda_1 L^2 (R_t^t+R_r^r)\nonumber \\&+&\frac{L^4}{12}[\frac{3}{2}\mu_1
(R^{rm}R_{rm}+R^{tm}R_{tm})+\mu_2(R_{mn}R^{mn}+R(R^r_r+R^t_t))+3\mu_3 R^2]\bigg{)}\,,\nonumber \\
\end{eqnarray}
so that using  eq. (\ref{bhmet}), it can be easily verified that  eq. (\ref{ent}) is reproduced. The thing to note here is that for eqs. (\ref{bhmet},\ref{bhmetp}), the
expression in the brackets in equation (\ref{wald}) is actually independent of $r$! In the context of the
observation in \cite{ms} this is particularly interesting since there a similar expression evaluated at the
boundary is related to a central charge.

We note here that if we evaluated the expression in the brackets in (\ref{wald}) for the metric ansatz in (\ref{ans1}) then we would get
\be\label{wald2}
\frac{1}{2}g_{rr}g_{tt}\frac{\partial {\mathcal L}}{\partial R_{rtrt}}=1+2 L^2 \lambda_1 A'(r)^2+L^4\mu_3 A'(r)^4=\lp c(r) A'(r)\,,
\ee
where ${\mathcal L}$ is the lagrangian.
This is quite remarkable since it shows that there is a connection between Wald's formula and the $c$-function! A similar observation is reported in \cite{ms} for higher dimensions. This makes it tempting to conjecture that there is a Wald like formula for $c(r)$ in any dimensions. In the context of two derivative theories, a covariant formula for $c(r)$ has been given in \cite{sahakian}. The fact that the spacetime central charge in AdS$_3$ could be extracted using a $c$-extremisation and its connection with the Wald formula was shown in \cite{kl}.
%In 1+1 dimension we have
%\be
%\frac{s}{T}=2\pi^2 c\,.
%\ee
%In \cite{ms} the generalization of this in 3+1 dimensions is
%\be
%\frac{s}{T^3}=2\pi^2 a\,,
%\ee
%where $a$ is the trace anomaly associated with the Euler density. {\bf Evaluate Wald on general backgd. Does it give $c(r)$ or something prop?}

\section{Unitarity and central charge}
For the NMG theory in the context of AdS/CFT that we have been considering, it has been shown that unitarity in
the bulk corresponds to negative CFT central charge. We wish to investigate the issue of unitarity in the
context of the extended NMG. In \cite{mgads}, the issue of unitarity was addressed by rewriting the NMG action in terms of a Pauli-Fierz lagrangian. In our case, we do not have this technology at our disposal and will use a somewhat indirect argument.
We begin with the ansatz
\be
ds^2=e^{2r/\tilde L} (-dt^2+dx^2)+dr^2+h_{ab}dx^a dx^b\,.
\ee
We choose the gauge $h_{rx}=h_{rt}=h_{rr}=0$. Then the linearised equations of motion for $h_{tt}, h_{xx} h_{tx}$ allow for the solution
$h_{tt}-h_{xx}=H(r), h_{tt}+h_{xx}=\tilde H(r), h_{tx}=G(r)$ where
\be
H''(r)-\frac{2}{\tilde L} H'(r)=0\,,
\ee
while both $G,\tilde H$ satisfy the equation
\be
2(1+2\fin \lambda+f^2_\infty \mu_3)F'-\tilde L[(1+18\fin \lambda_1+17f^2_\infty \mu_3)F''+4\tilde L(\fin\lambda_1+f^2_\infty\mu_3)(-4 F'''+\tilde L F'''')]=0\,,
\ee
which can be rewritten as
\be \label{factor}
(\partial_r^2-\frac{2}{\tilde L}\partial_r)(\partial_r^2-\frac{2}{\tilde L}\partial_r-\frac{\lp c \fin}{4 \tilde L^3(1-\fin)})F=0\,,
\ee
where $F$ is either $\tilde H$ or $G$. Thus while $H$ is massless both $\tilde H, G$ allow for massive modes. This is the key step. Note here that the way we have written the above equation, there is no explicit dependence on $\lambda_1,\mu_3$ in the equations of motion.
Making a change of variables $e^{2r/\tilde L}=R/\tilde L$, the massive mode satisfies
\be
(R^2\partial_R^2 -\frac{\lp c \fin}{16\tilde L (1-\fin)})F=0\,,
\ee
which has the solution
\be
F\sim R^{\frac{1}{2}(1\pm \sqrt{1+\frac{c \fin \lp}{4 \tilde L(1-\fin)}})}\,.
\ee
The fluctuation should go like $R^a$ with $a\leq 1$ (as otherwise it will spoil the asymptotics) and should be well behaved at the origin $R=0$ so that we must have
\be
\frac{c \fin \lp}{4 \tilde L(1-\fin)}\leq 0\,.
\ee
This means either $c\geq 0, \fin >1$ or $c\leq 0, \fin<1$.  The inequalities are valid both for the NMG and the extended NMG. The prefactor for the kinetic term for the massive modes in both cases is proportional to  $(1-\fin)$. As such only for the special case when $\fin=1$ or equivalently $- \lambda_1=\mu_3$ will the ghost problem be ameliorated since in this case the equations of motion work out to be two derivative. More specifically, it can be shown that if $\mu_3<1$ then the theory is unitary in this case. Note that it was crucial for the cubic term to be present for this possibility to be realized. A more general analysis involving a general coordinate dependence for the fluctuations was performed in \cite{paulos} and corroborates this finding. If $\fin\neq 1$, for the massive modes, to be compatible\footnote{In \cite{mgads} the condition for unitarity for the NMG theory ($\mu_3=0$) around AdS is $\frac{\Lambda-2m^2}{4 m^2}\geq 0$. In our notation $m^2=1/(4\lambda_1 L^2),\Lambda=-\fin/L^2$ so that we get $1+2\lambda_1 \fin \leq 0$ in other words $c\leq 0$. Also note that the kinetic term for the massless non-propagating gravity is positive only if $c\geq 0$. It is this piece that leads to the two point function for stress tensors in the CFT.} with the analysis for unitarity in the NMG in \cite{mgads}, $c\leq 0$ should be chosen. It is only for this choice that one can have unitary massive spin-2 modes in the bulk.
%Another way to see this is as follows. Eq. (\ref{factor}) can be written as
%\be
%(\partial_r^2-\frac{2}{\tilde L}\partial_r)((1-\fin)\partial_r^2-\frac{2(1-\fin)}{\tilde L}\partial_r-\frac{\lp c \fin}{4 \tilde L^3})F=0\,.
%\ee
%Now when we set $\lambda_1=\mu_3=0$, then we get the eom \cite{mr} to be $-c (\partial_r^2-\frac{2}{\tilde L}\partial_r)F=0$. The test for unitarity in this case is $c\geq 0$. In the same vein, in the presence of $\lambda_1,\mu_3$, i.e., when $1-\fin\neq 0$, we expect that the test for unitarity will be that the coefficient of $\partial_r^2$ in the above equation will be negative. This leads to $c\leq 0, \fin<1$ to be singled out as the condition for unitarity.
{\it Thus we conclude that bulk unitarity for the massive modes implies that the CFT central charge is negative or zero.}

\section{Conclusion}
In this paper we have showed that the new massive gravity theory admits for a $c$-theorem. Turning the argument around we have showed that demanding the existence of a $c$-function similar to that in GB gravity in higher dimensions \cite{ms}, the NMG emerges at the four derivative order. We have constructed a set of six derivative terms involving the curvature invariants which admit a similar holographic $c$-theorem. One motivation to consider such an extension is to see what this has in common with the NMG. We find the following common features
\begin{itemize}
\item Lifshitz solutions are allowed as shown in appendix B.
\item Like the NMG, the extended theory admits for exact black hole solutions in AdS. For a special choice of the parameters, extremal solutions are allowed.
\item The $c$-function can be written in terms of the Wald formula as in eq. (\ref{wald2}).
\item For fluctuations depending only on the radial coordinate, the equations of motion are four derivative. A nice factorization property \cite{mgads2} as shown in (\ref{factor}) is at work for both theories.
\item Bulk unitarity is still in conflict with the positivity of the boundary central charge for generic choice of parameters. For the case when $\lambda_1=-\mu_3$, the linearised equations of motion become two derivative. For the NMG, at the special point where the central charge vanishes it was shown in \cite{gh} that the dual CFT (if it exists) becomes a logarithmic CFT. The conflict between bulk unitarity vs positive CFT central charge is less severe in this case; for example there are no negative mass black holes. It is likely that the story is the same for the extended version of NMG.
\end{itemize}

Apriori, our method which led to constraining the higher order lagrangians is not guaranteed
to work order by order. However, we have demonstrated that upto eight derivative order (see appendix A), one can construct a
holographic $c$-theorem as dictated by 1+1 dimensional dual CFTs. It will be interesting to study these theories
further. It will be very interesting to see which of the above features are common to any theory. For example using the methods in this paper can some interesting inequality for the dynamical exponent in Lifshitz metrics be derived? It will also be useful to consider finite coupling effects or higher derivative corrections to holographic quantum liquids in 1+1 dimensions \cite{hs}.

\section*{Acknowledgments}
I thank A. Buchel, D. Jatkar, R. Myers for discussions and comments on the draft. I also thank P. Townsend and O. Hohm for comments on the draft and useful correspondence.
I thank HRI for kind hospitality during the initial stages of this work. Research at Perimeter Institute is supported by the Government of Canada through Industry Canada and by the Province of
Ontario through the Ministry of Research \& Innovation.
\begin{appendix}
\section{Extending to eight derivative gravity}
The original NMG theory can be rewritten in terms of a Pauli-Fierz lagrangian. For this the starting point is
the observation that the NMG theory is equivalent to
\be\label{pf}
I_{eq}=\int d^3 x
\sqrt{-g}\left(R+\frac{2}{L^2}+f^{\mu\nu}\G_{\mu\nu}+\alpha(f^{\mu\nu}f_{\mu\nu}-f^2)\right)\,,
\ee
where $\G_{\mu\nu}=R_{\mu\nu}-1/2 R g_{\mu\nu}$, $f=f_{\mu\nu}g^{\mu\nu}$ and $\alpha=-1/(16\lambda_2 L^2)$. One obvious way to extend
to higher derivative lagrangians is to consider replacing $\G_{\mu\nu}$ by adding four derivative terms. However,
it is easy to see that this will generically lead to eight derivative terms in the lagrangian after integrating out $f_{\mu\nu}$.
Motivated by this we extend our method of constructing higher derivative lagrangians to eight derivative order.
The terms we will consider are
\be
\R_6= L^6 [\nu_1 R_{\mu}^\nu R_\nu^\rho R_\rho^\lambda R_\lambda^\mu+\nu_2 (R_{\mu\nu}R^{\mu\nu})^2+\nu_3
R_{\mu\nu}R^{\mu\nu}R^2+ \nu_4 R^4+\nu_5 R_\mu^\nu R_\nu^\rho R_\rho^\mu R]\,,
\ee
using which it can be shown that choosing
\be
\nu_1=-\frac{41}{20}\nu-6 \nu_4\,,\quad \nu_2=\frac{21}{8} \nu+3 \nu_4\,,\quad \nu_3=-\frac{17}{20}\nu-6\nu_4\,,\quad
\nu_5=\frac{3}{5}\nu+8\nu_4\,,
\ee
will lead to
\be
c(r)=\frac{1}{\lp A'(r)}(1+2\lambda_2 L^2 A'^2+\mu_3 L^4 A'^4+\frac{4}{11}\nu L^6 A'^6)\,,
\ee
being the $c$-function with $c'(r)\geq 0$. All the properties found before extend to this theory. We now replace
\be
\G_{\mu\nu}=R_{\mu\nu}-1/2 g_{\mu\nu}R+\beta_1 R_{\mu\rho}R^{\rho}_\nu+\beta_2 R^2 g_{\mu \nu}+\beta_3 R R_{\mu
\nu}+ \beta_4 R_{\lambda\rho}R^{\lambda\rho}g_{\mu\nu}\,,
\ee
with
\begin{eqnarray}
\beta_1&=&-\frac{32}{51} \alpha \mu_3\,,\quad \beta_2=-\frac{10}{51}\alpha\mu_3\,,\quad \beta_3=-\frac{10}{51} \alpha \mu_3\,,\quad
\beta_4=\frac{16}{51} \alpha \mu_3\,,\\
\nu_4&=& \frac{11}{578} \alpha \mu_3^2\,,\quad \nu=-\frac{20}{2601} \alpha \mu_3^2\,,
\end{eqnarray}
will lead to an action of the form (\ref{pf}).
\section{Lifshitz type solutions}
It was shown in \cite{amsv} that in a general higher derivative gravity theory, Lifshitz solutions will emerge. The same statement is true in the extended NMG theory and has been studied in related contexts in \cite{lif}. Making the ansatz
\be
ds^2=-\frac{r^{2+\alpha}}{L^{2+\alpha}}dt^2+\frac{r^2}{L^2} dx^2+ \frac{L^2}{\fin}\frac{dr^2}{r^2}\,,
\ee
the equations of motion lead to the conditions
\begin{eqnarray}
0&=&32-16\fin(2+\alpha)-2f^2_\infty \lambda_1 (\alpha^4-4\alpha^2-16\alpha-16)\nonumber \\ &&~~~~-\mu_3\fin^3(\alpha^6+7\alpha^5-2\alpha^4-68\alpha^3-88\alpha^2-48\alpha-32)\,,\\
0 &=& 32-32\fin -2f^2_\infty \lambda_1 (\alpha^4+4\alpha^3-12\alpha^2-32\alpha-16)\nonumber \\&&~~~~-\mu_3f^3_\infty(\alpha^6+10\alpha^5+26\alpha^4-16\alpha^3-104\alpha^2-64\alpha-32)\,.
\end{eqnarray}
Note that $\alpha=0$ is always a solution for any value of $\lambda_1,\mu_3$. We are interested in $\alpha\neq 0$. First consider the case $\mu_3=0$. For this we can solve for $\fin$ and $\alpha$ to get the following plots.
\FIGURE[h]{\begin{tabular}{cc}
\includegraphics[width=0.45 \textwidth]{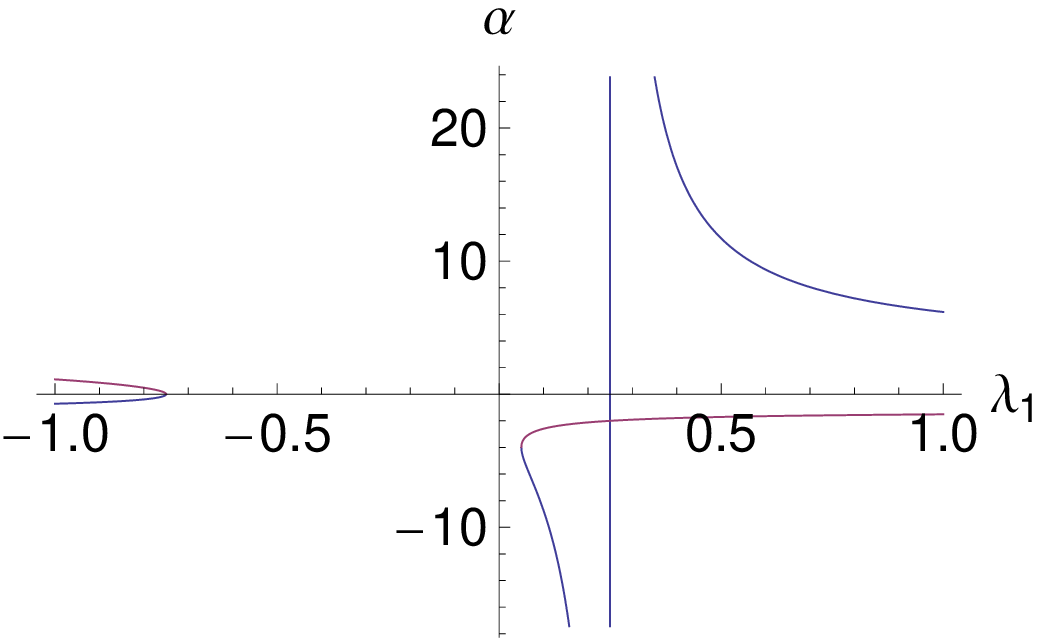}&
\includegraphics[width=0.45 \textwidth]{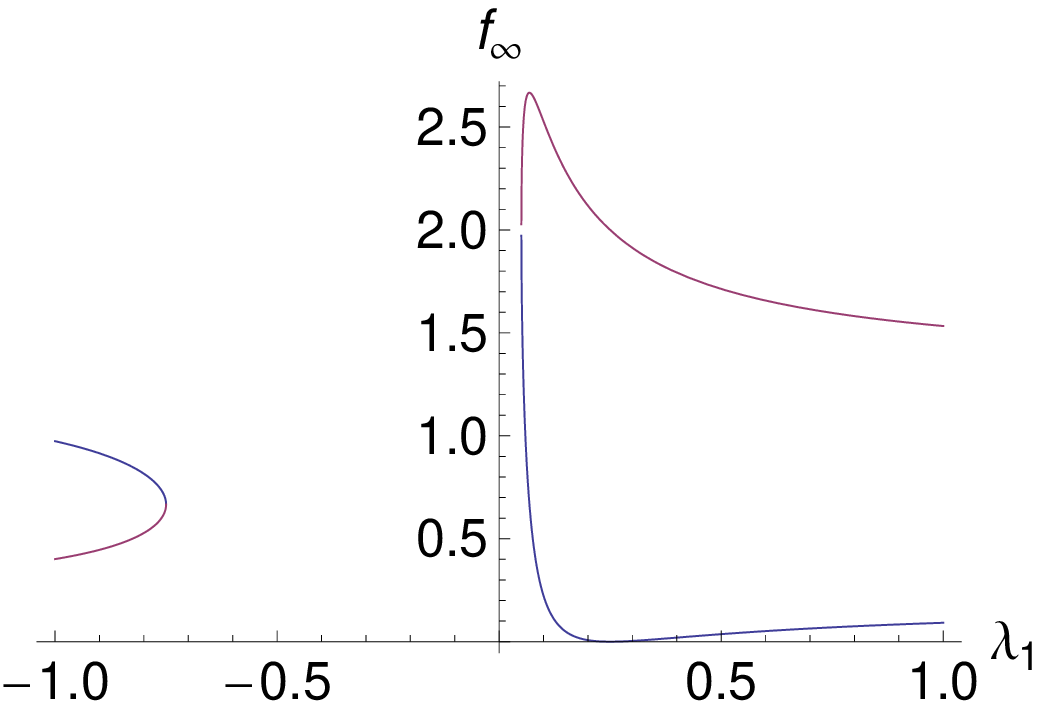}\end{tabular}
\caption{The plot on the left shows $\alpha$ vs $\lambda_1$ when $\mu_3=0$ while that on the right shows $\fin$ vs $\lambda_1$ when $\mu_3=0$. } \label{plotapp1}}
When $\mu_3\neq 0$ we expect to have a 3-d plot. We can for example set $\alpha=2$ to see a cross-section of that plot. This leads to the plots in fig. (\ref{plotapp2}).
\FIGURE[h]{\begin{tabular}{cc}
\includegraphics[width=0.45 \textwidth]{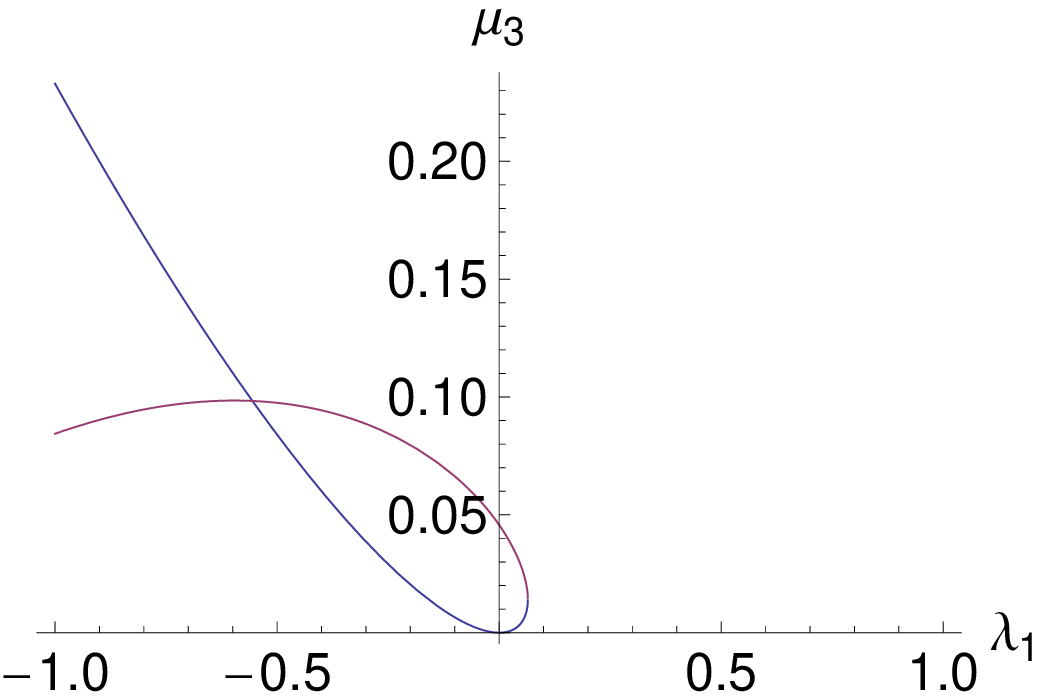}&
\includegraphics[width=0.45 \textwidth]{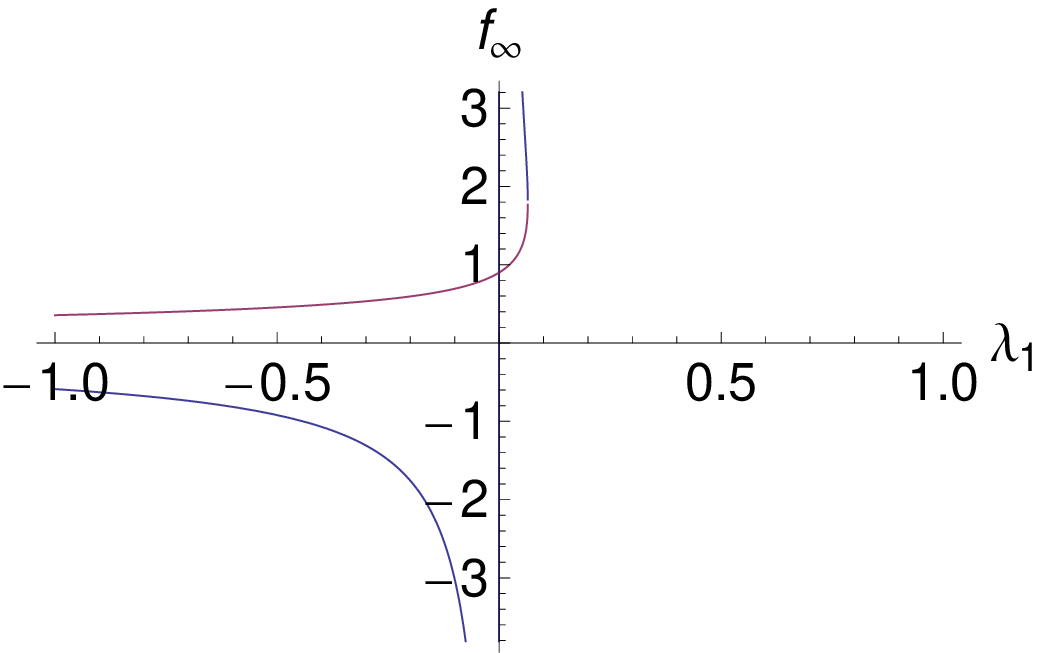}\end{tabular}
\caption{The plot on the left shows $\mu_3$ vs $\lambda_1$ when $\alpha=2$ while that on the right shows $\fin$ vs $\lambda_1$ when $\alpha=2$. } \label{plotapp2}}
Note that the blue branch in fig. (\ref{plotapp2}) has negative values of $\fin$.

\section{More on the $c$ function}
Consider the case where we have included four derivative terms. If we demand that the $c$-function is made of $A'$ and derivatives, then we can prove that the only way to make $c'$ fulfill eq. (\ref{only}) is to set  $\lambda_2=-3/8\lambda_1$. It is obvious that the most general form for $c$ is \be
c=\frac{1}{\lp A'}(1+F[A',A'',A'''])\,.
\ee
We cannot have $A''''$ in $c$ since then $c'$ will involve five derivatives so that using the equations of motion will involve $\partial_r(T^t_t-T^r_r)$ and we do not want to impose a constraint on this. Then
\be
c'=-\frac{1}{\lp A'^2}\left(A'' (1+F)- A'(\partial_{A'}F A''+ \partial_{A''}F A'''+\partial_{A'''}F A'''')\right)\,.
\ee
The term in brackets should be proportional to the equations of motion. Comparing the four derivative term gives
\be
\partial_{A'''}F A' =4 L^2 (8\lambda_2+3\lambda_1)\,,
\ee
so that
\be
F= 4 L^2 (8\lambda_2+3\lambda_1)\frac{A'''}{A'}+\tilde F[A',A'']\,.
\ee
Now focus on $A'''$. This will lead to
\be
8 L^2 (8\lambda_2+3\lambda_1)\frac{ A''}{A'}-A'\partial_{A''}\tilde F =-8 L^2 (8\lambda_2+3\lambda_1)A'\,,
\ee
which leads to
\be
\tilde F=4 L^2 (8\lambda_2+3\lambda_1) \frac{(A'')^2}{A'^2}+8 L^2 (8\lambda_2+3\lambda_1) A'' +\hat F[A']\,.
\ee
Next consider the terms proportional to $A''$. There are at this stage in the expression for $c'$ the term
\be
-12 L^2(3 \lambda_1+8 \lambda_2)\frac{(A'')^3}{(A')^4} \,,
\ee
which has no analogue in the equations of motion. Thus this has to be set to zero leading to $\lambda_2=-3/8\lambda_1$. This completes the proof.
\end{appendix}


\begin{thebibliography}{99}

\bibitem{nmg}
  E.~A.~Bergshoeff, O.~Hohm and P.~K.~Townsend,
  ``Massive Gravity in Three Dimensions,''
  Phys.\ Rev.\ Lett.\  {\bf 102}, 201301 (2009)
  [arXiv:0901.1766 [hep-th]].

\bibitem{nmg2}
  E.~Bergshoeff, O.~Hohm and P.~Townsend,
  ``On massive gravitons in 2+1 dimensions,''
  arXiv:0912.2944 [hep-th].
\bibitem{nmg3}
  R.~Andringa, E.~A.~Bergshoeff, M.~de Roo, O.~Hohm, E.~Sezgin and P.~K.~Townsend,
  ``Massive 3D Supergravity,''
  Class.\ Quant.\ Grav.\  {\bf 27}, 025010 (2010)
  [arXiv:0907.4658 [hep-th]].

\bibitem{many}
  O.~Hohm and E.~Tonni,
  ``A boundary stress tensor for higher-derivative gravity in AdS and Lifshitz
  backgrounds,''
  arXiv:1001.3598 [hep-th].
  %K.~Skenderis, M.~Taylor and B.~C.~van Rees,
  %``Topologically Massive Gravity and the AdS/CFT Correspondence,''
  %JHEP {\bf 0909}, 045 (2009)
  %[arXiv:0906.4926 [hep-th]].

\bibitem{unit}
  S.~Deser,
  ``Ghost-free, finite, fourth order D=3 (alas) gravity,''
  Phys.\ Rev.\ Lett.\  {\bf 103}, 101302 (2009)
  [arXiv:0904.4473 [hep-th]].

\bibitem{mgads}
  E.~A.~Bergshoeff, O.~Hohm and P.~K.~Townsend,
  ``More on Massive 3D Gravity,''
  Phys.\ Rev.\  D {\bf 79}, 124042 (2009)
  [arXiv:0905.1259 [hep-th]].



\bibitem{mgads2}
  Y.~Liu and Y.~W.~Sun,
  ``Note on New Massive Gravity in $AdS_3$,''
  JHEP {\bf 0904}, 106 (2009)
  [arXiv:0903.0536 [hep-th]].

\bibitem{mgads3}
  Y.~Liu and Y.~W.~Sun,
  ``Consistent Boundary Conditions for New Massive Gravity in $AdS_3$,''
  JHEP {\bf 0905}, 039 (2009)
  [arXiv:0903.2933 [hep-th]].


\bibitem{gbhd}
  A.~Buchel, J.~Escobedo, R.~C.~Myers, M.~F.~Paulos, A.~Sinha and M.~Smolkin,
  ``Holographic GB gravity in arbitrary dimensions,''
  arXiv:0911.4257 [hep-th].

\bibitem{gbhd2}
  X.~O.~Camanho and J.~D.~Edelstein,
  ``Causality constraints in AdS/CFT from conformal collider physics and
  Gauss-Bonnet gravity,''
  arXiv:0911.3160 [hep-th].

\bibitem{gbhd3}
  J.~de Boer, M.~Kulaxizi and A.~Parnachev,
  ``$AdS_7/CFT_6$, Gauss-Bonnet Gravity, and Viscosity Bound,''
  arXiv:0910.5347 [hep-th].


\bibitem{ms} 
  R.~C.~Myers and A.~Sinha,
  ``Seeing a c-theorem with holography,''
  arXiv:1006.1263 [hep-th].

\bibitem{cth}
  D.~Z.~Freedman, S.~S.~Gubser, K.~Pilch and N.~P.~Warner,
  ``Renormalization group flows from holography supersymmetry and a
  c-theorem,''
  Adv.\ Theor.\ Math.\ Phys.\  {\bf 3}, 363 (1999)
  [arXiv:hep-th/9904017].

\bibitem{no}
  M.~Nakasone and I.~Oda,
  ``On Unitarity of Massive Gravity in Three Dimensions,''
  Prog.\ Theor.\ Phys.\  {\bf 121}, 1389 (2009)
  [arXiv:0902.3531 [hep-th]].
  I.~Gullu and B.~Tekin,
  ``Massive Higher Derivative Gravity in D-dimensional Anti-de Sitter
  Spacetimes,''
  Phys.\ Rev.\  D {\bf 80}, 064033 (2009)
  [arXiv:0906.0102 [hep-th]].


\bibitem{mr} R.~C.~Myers and B.~Robinson, ``Black Holes in Quasi-topological Gravity,''
  arXiv:1003.5357 [gr-qc].


\bibitem{tmg}
  S.~Deser, R.~Jackiw and S.~Templeton,
  ``Topologically massive gauge theories,''
  Annals Phys.\  {\bf 140}, 372 (1982).
W.~Li, W.~Song and A.~Strominger,
  ``Chiral Gravity in Three Dimensions,''
  JHEP {\bf 0804}, 082 (2008)
  [arXiv:0801.4566 [hep-th]].


\bibitem{tmgcth}
  K.~Hotta, Y.~Hyakutake, T.~Kubota, T.~Nishinaka and H.~Tanida,
  ``Left-Right Asymmetric Holographic RG Flow with Gravitational Chern-Simons
  Term,''
  Phys.\ Lett.\  B {\bf 680}, 279 (2009)
  [arXiv:0906.1255 [hep-th]].






\bibitem{mps}R.~C.~Myers, M.~F.~Paulos and A.~Sinha,
  ``Holographic studies of quasi-topological gravity,''
  arXiv:1004.2055 [hep-th].

\bibitem{ejm}
  R.~Emparan, C.~V.~Johnson and R.~C.~Myers,
  ``Surface terms as counterterms in the AdS/CFT correspondence,''
  Phys.\ Rev.\  D {\bf 60}, 104001 (1999)
  [arXiv:hep-th/9903238].





\bibitem{amsv}
  A.~Adams, A.~Maloney, A.~Sinha and S.~E.~Vazquez,
  ``1/N Effects in Non-Relativistic Gauge-Gravity Duality,''
  JHEP {\bf 0903}, 097 (2009)
  [arXiv:0812.0166 [hep-th]].

\bibitem{hs}
  L.~Y.~Hung and A.~Sinha,
  ``Holographic quantum liquids in 1+1 dimensions,''
  JHEP {\bf 1001}, 114 (2010)
  [arXiv:0909.3526 [hep-th]].

\bibitem{lif}
R.~G.~Cai, Y.~Liu and Y.~W.~Sun,
  ``On the z=4 Horava-Lifshitz Gravity,''
  JHEP {\bf 0906}, 010 (2009)
  [arXiv:0904.4104 [hep-th]].
R.~G.~Cai, Y.~Liu and Y.~W.~Sun,
  ``A Lifshitz Black Hole in Four Dimensional $R^2$ Gravity,''
  JHEP {\bf 0910}, 080 (2009)
  [arXiv:0909.2807 [hep-th]]. E.~Ayon-Beato, A.~Garbarz, G.~Giribet and M.~Hassaine,
  ``Lifshitz Black Hole in Three Dimensions,''
  Phys.\ Rev.\  D {\bf 80}, 104029 (2009)
  [arXiv:0909.1347 [hep-th]]. Y.~S.~Myung, Y.~W.~Kim and Y.~J.~Park,
  ``Dilaton gravity approach to three dimensional Lifshitz black hole,''
  arXiv:0910.4428 [hep-th].
E.~Ayon-Beato, A.~Garbarz, G.~Giribet and M.~Hassaine,
  ``Analytic Lifshitz black holes in higher dimensions,''
  arXiv:1001.2361 [hep-th].






\bibitem{sahakian}
  V.~Sahakian,
  ``Holography, a covariant c-function and the geometry of the  renormalization
  group,''
  Phys.\ Rev.\  D {\bf 62}, 126011 (2000)
  [arXiv:hep-th/9910099].

\bibitem{kl}
  P.~Kraus and F.~Larsen,
  ``Microscopic Black Hole Entropy in Theories with Higher Derivatives,''
  JHEP {\bf 0509}, 034 (2005)
  [arXiv:hep-th/0506176].


\bibitem{gh}
  D.~Grumiller and O.~Hohm,
  ``$AdS_3/LCFT_2$ - Correlators in New Massive Gravity,''
  arXiv:0911.4274 [hep-th].

\bibitem{tekin}
  I.~Gullu, T.~C.~Sisman and B.~Tekin,
  ``Born-Infeld extension of new massive gravity,''
  arXiv:1003.3935 [hep-th].

\bibitem{Oliva:2009ip}
  J.~Oliva, D.~Tempo and R.~Troncoso,
  ``Three-dimensional black holes, gravitational solitons, kinks and wormholes
  for BHT masive gravity,''
  JHEP {\bf 0907}, 011 (2009)
  [arXiv:0905.1545 [hep-th]]. G.~Giribet, J.~Oliva, D.~Tempo and R.~Troncoso,
  ``Microscopic entropy of the three-dimensional rotating black hole of BHT
  massive gravity,''
  Phys.\ Rev.\  D {\bf 80}, 124046 (2009)
  [arXiv:0909.2564 [hep-th]].

\bibitem{wald}
 R.~M.~Wald,
  ``Black hole entropy is the Noether charge,''
  Phys.\ Rev.\  D {\bf 48}, 3427 (1993)
  [arXiv:gr-qc/9307038].

\bibitem{paulos}
  M.~F.~Paulos,
  ``New massive gravity, extended,''
  arXiv:1005.1646 [hep-th].




\end{thebibliography}
\end{document}